\documentstyle[emulateapj,psfig]{article}
\slugcomment{The Astrophysical Journal (Letters), in press}
\lefthead{Brainerd}
\righthead{Anisotropic Distribution of SDSS Satellite Galaxies}

\begin{document}

\title{Anisotropic Distribution of SDSS Satellite Galaxies: \\
Planar (not
Polar) Alignment}

\author{Tereasa G.\ Brainerd}
\affil{Boston University, Institute for Astrophysical Research, 
725 Commonwealth Ave., 
Boston, MA 02215}
\affil{brainerd@bu.edu}

\begin{abstract}
The distribution of satellite galaxies relative to isolated
host galaxies in the
Sloan Digital Sky Survey (SDSS) is investigated.  
Host--satellite systems are selected
using three different methods, yielding samples of $\sim 3300$,
$\sim 1600$, and $\sim 950$ satellites.  
In the plane of the sky, the distributions of all
three samples show highly significant 
deviations
from circular symmetry 
($> 99.99$\%,
$> 99.99$\%, and 99.79\% confidence
levels, respectively), and
the degree of anisotropy 
is a strong function of the projected radius,
$r_p$,
at which the satellites are found.  For 
$r_p \lesssim 100$~kpc, the SDSS satellites are aligned preferentially with
the major axes of the hosts.  This is in stark contrast to the Holmberg
effect, in which satellites are aligned with the minor axes of
host galaxies.  The degree of anisotropy in the distribution of 
the SDSS satellites decreases with $r_p$ and is
consistent with an isotropic distribution at of order the 1-$\sigma$ level for
$250~{\rm kpc} \lesssim r_p \lesssim 500$~kpc.
\end{abstract}


\section{Introduction}

The distribution of satellite galaxies relative to their
host galaxies may hold important clues to the 
formation history of large galaxies.  It should reflect the time
scales for dwarf galaxy evolution and orbital decay, as well the nature
of the host's dark matter halo  and the
manner in which the host  accretes mass.
On scales of order the virial radius,
$r_{200}$,
the halos of large galaxies in cold dark matter (CDM) universes
are flattened with a mean projected ellipticity
of order 0.3 (e.g., Dubinski \& Carlberg 1991; Warren et al.\ 1992; Jing
\& Suto 2002).  On small scales, $r << r_{200}$, gas cooling
makes the halos somewhat rounder, but on scales $r \sim r_{200}$
baryons do not affect the shapes of the halos very much (e.g., Kazantzidis
et al.\ 2004).
Evidence for non--spherical halos 
includes dynamics of polar ring galaxies, geometry
of X-ray isophotes, flaring of HI gas in spirals, evolution
of gaseous warps, kinematics of Population II stars in our
own Galaxy, and gravitational lensing (se, e.g.,  Sackett 1999;
Keeton et al.\ 1998; Maller et 
al.\ 2000; Hoekstra et al.\ 2004).  However, Ibata et al.\
(2001) concluded a nearly spherical halo for the Milky Way on the
basis of tidal stellar streams.

If the halos of large galaxies are flattened, 
the distribution of their satellites may
reflect the flattening of the halo and, hence, show
some preference for alignment with the major axis of the halo mass 
distribution, as projected on the sky.  Assuming the light
and mass of a large galaxy are aligned reasonably well,
one then expects a preference for satellites
to be aligned with the major axes of the images of their hosts. 
However, this has not generally been observed.
Holmberg (1969) investigated the distribution of satellites
relative to disk galaxy hosts and, for projected radii
$r_p \lesssim 50$~kpc, found a tendency for the satellites to be located
preferentially close to the minor axes of the hosts.  In addition,
the 11 brightest satellites of the Milky Way form a flattened structure
that is oriented roughly perpendicular to the disk (e.g., Lynden--Bell
1982; Majewski 1994).  Conversely,
Valtonen et al.\ (1978) found a tendency for compact satellites
to be aligned with the major axes of disk hosts.
Hawley \& Peebles (1975), Sharp et al.\ (1979), and
MacGillivray et al.\ (1982), suggested that any anisotropy in the
distribution of satellite galaxies was
at most rather small, and perhaps non--existent.
Zaritsky et al.\ (1997),
hereafter ZSFW,
studied the distribution of satellites around spiral hosts
and found evidence for an alignment of the
satellites along the host minor axes for 
$300~{\rm kpc} \lesssim r_p \lesssim 500~{\rm kpc}$.
Sales \& Lambas (2004), hereafter SL,
used a set of 1498 host galaxies
with 3079 satellites from the Two Degree Field Galaxy Redshift
Survey (2dFGRS; Colless et al.\ 2001, 2003) to investigate the distribution of
satellites and
found a preference for the satellites to be aligned with the
minor axes of the hosts, but only for a very specific
subsample of host--satellite pairs.

The theory regarding the Holmberg
effect is equally as murky as the observations.  ZSFW
were unable to recover their observed anisotropy from
CDM simulations.  Abadi et al.\ (2003)
suggested that the Holmberg effect could be caused by the cumulative
effects of accretion of satellites by the host.  
Pe\~narrubia
et al.\ (2002) suggested that the Holmberg effect orginates with
the fact that polar orbits 
inside a massive,
flattened halo do not decay as quickly as planar orbits.
However, Knebe et al.\ (2004) found that the 
orbits of satellites
of primary galaxies in cluster environments were located preferentially
within a cone of opening angle 40$^\circ$ (i.e., planar alignment,
not polar).  Since the structure of CDM halos is
largely independent of the halo mass scale (e.g., Moore et al.\  1999), the
implication of this result is a
preference for the satellites of
isolated galaxies to be aligned with the major axes of the host
halos. Indeed, the most
recent numerical work seems to show that this the case (e.g., Agustsson
\& Brainerd 2005, hereafter AB; Libeskind et al.\ 2005; Zentner et al.\ 2005).

The organization of this {\it Letter} is as follows.
The distribution of satellite galaxies 
relative to host galaxies
in the third data release (DR3) of the SDSS
(Abazajian et al.\ 2005; Fukugita et al.\ 1996; Hogg et al.\ 2001;
Smith et al.\ 2002; Strauss et al.\ 2002; York et al.\ 2002) 
is shown in \S2 and is compared to an isotropic
distribution.  The results obtained in \S2 are
compared to other recent observational and numerical results in \S3.
Throughout, the null hypothesis is that satellites are spherically
distributed about their hosts, giving rise to a circularly--symmetric
distribution on the
sky.  A rotation of
a spherically--symmetric
distribution through any combination of Euler angles always 
gives rise to a 
2-d distribution that is circularly--symmetric and, hence, any
deviation in the satellite distribution from pure circular symmetry
cannot be caused simply by projection and/or rotation effects.
Cosmological 
parameters $\Omega_0 = 0.3$, $\Lambda_0 = 0.7$, and 
$H_0 = 70$~km~sec$^{-1}$~Mpc~$^{-1}$ are adopted below.  

\section{Host--Satellite Selection and the Distribution of Satellites}

The $r'$ band photometric data from the SDSS DR3 were used to define
the apparent magnitudes of the galaxies, their position angles, and
the shapes of the equivalent image ellipses. 
Only galaxies for which the SDSS redshift confidence
parameter, zconf, is greater than 0.9 were used.
Three methods were adopted in order
to select isolated host galaxies
and their satellites.
``Sample 1'' was selected using the criteria of 
SL:
[1] Hosts are at least 2.5 times brighter than any other galaxy
within a projected radius of $r_p < 700$~kpc and a relative
radial velocity difference of $|dv| < 1000$~km~sec$^{-1}$.
[2] Satellites are at least 6.25 times fainter than their host,
are found within $r_p < 500$~kpc,
and the host--satellite
velocity difference is $|dv| < 500$~km~sec$^{-1}$.
``Sample 2'' was selected using the criteria of McKay et al.\ (2002) in
their study of the dynamics of satellite galaxies in the SDSS, with
the exception that here a slightly smaller maximum $r_p$ for
the satellites was used in order to be consistent with the other two samples:
[1] Hosts are
at least 2 times brighter than any other galaxy 
within 
$r_p < 2.86$~Mpc and 
$|dv| <  1000$~km~s$^{-1}$.
[2] Satellites are at least 4 times fainter
than their host, are found within
$r_p < 500$~kpc
(cf. $r_p < 714$~kpc in McKay et al.\ 2002), and
have $|dv| < 1000$~km~s$^{-1}$.
``Sample 3'' was selected using the criteria of Zaritsky et al.\ (1993) and
ZSFW:
[1] Hosts are at least 8 times brighter than any other
galaxy within $r_p < 500$~kpc and  
$|dv| < 1000$~km~sec$^{-1}$.
In addition, hosts are at least 2 times brighter than any
other galaxy within $r_p < 1$~Mpc and
$|dv| < 1000$~km~sec$^{-1}$.
[2] Satellites are at least 8 times fainter than their host,
are found within $r_p < 500$~kpc, and 
have $|dv| < 500$~km~sec$^{-1}$.

Additionally, in order to eliminate hosts which have a very large number
of satellites (and may be associated with cluster systems), the
luminosity of a host must be greater than
the sum of the luminosities of its satellites.
Finally, the samples were restricted to hosts
with ellipticity is $\epsilon \ge 0.2$ so that the position angles
of the major axes of the hosts are well determined.
The selection criteria yield
1973 hosts and 3292 satellites for sample 1,
913 hosts and 1575 satellites for sample 2, and 
575 hosts and 935 satellites for sample 3.  
The median redshifts of the hosts are $z_{\rm med} = 0.05$,
0.08, and 0.03, respectively. The 
number of satellites per host is shown
in Fig.\ 1, from which it is clear that the samples are dominated
by systems containing one or two satellites.  Also, although
the morphologies of all of the hosts are not known, a visual inspection
of images of the 50 brightest hosts (all of which appear in all 
3 samples) shows that 44 are spirals (mostly Sc or Sd),
4 are S0,
and 2 are Sd/Irr. 

%
%
\vspace*{0.2truecm}
\hbox{~}
\centerline{\psfig{file=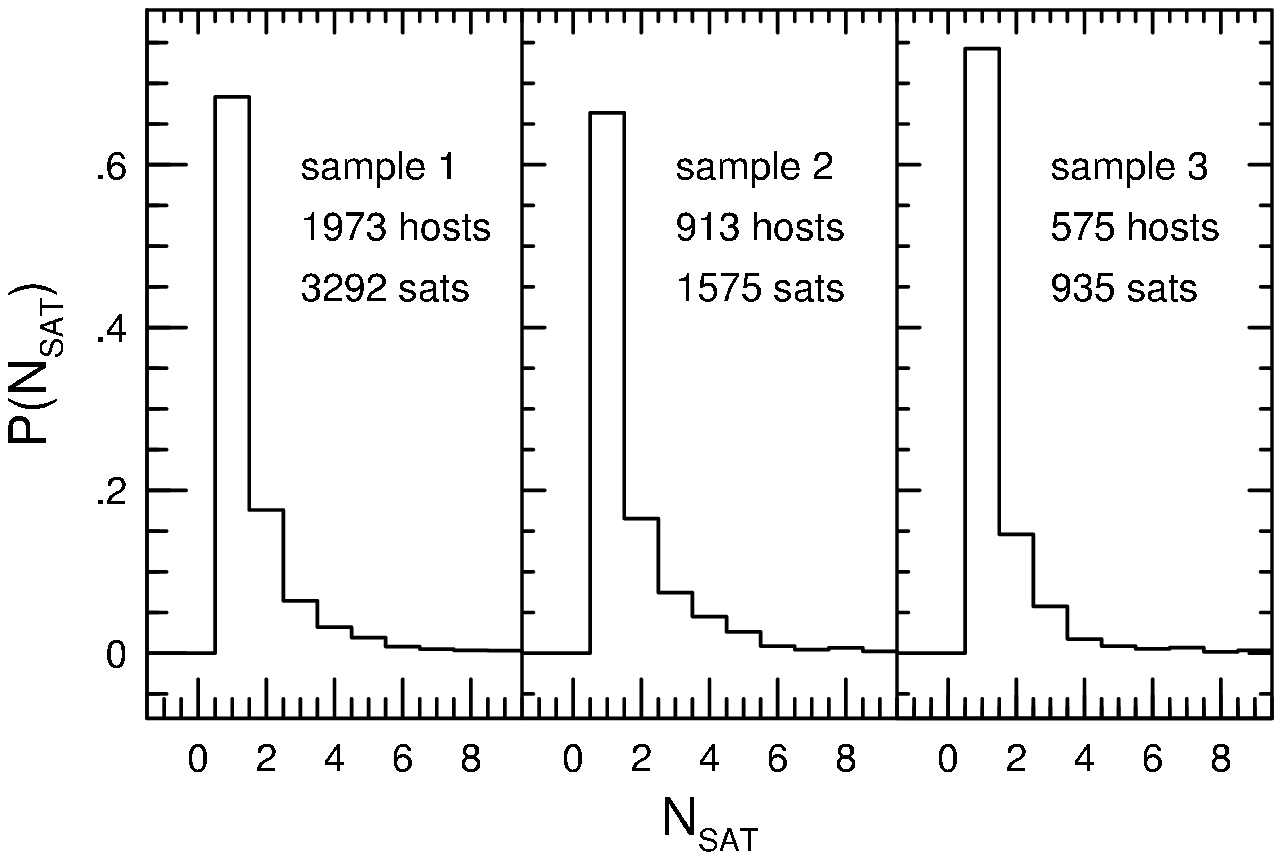,angle=0,width=3.25in}}
\noindent{\scriptsize
Figure~1.
Normalized probability distribution of the number of satellites
per host for each of the three samples of hosts and satellites
(see text for sample selection criteria).
\label{fig1}
\vspace*{0.2truecm}
}

The distribution of the satellites relative to their hosts was computed
by measuring the angle, $\phi$, between the major axis of the host and 
the direction vector on the sky that connects the centroids of the host
and its satellite.   Because we are
simply interested in investigating any preferential alignment
of the satellites with the principle axes of
the hosts, $\phi$ is restricted to the range 
$[0^\circ,90^\circ]$ (i.e., 
$\phi = 0^\circ$ indicates alignment with the host major axis
and $\phi = 90^\circ$ indicates alignment with the host minor
axis).  
Results of the analysis for 
all three samples are shown in Fig.\ 2.  The left panels show
the differential probability distribution, $P(\phi)$, and the right
panels show the cumulative probability distribution, $P(\phi \le 
\phi_{\rm MAX})$. 
The probabilities expected for an isotropic (i.e., circularly--symmetric)
distribution of satellites 
are shown by the dotted lines in all eight
panels of Fig.\ 2.  A $\chi^2$ test was applied to $P(\phi)$,
and an isotropic distribution of 
satellites was rejected for all three samples (see Fig.\ 2).
The mean value of $\phi$,
$\left< \phi \right>$, 
was also computed for each sample of satellites
and excellent agreement amongst the samples is found.  The mean value of
$\phi$ is considerably less than the expected value for an isotropic
distribution (i.e., $\left< \phi \right> = 45^\circ$), and differs formally
from the isotropic distribution by 6.4-$\sigma$, 
4.6-$\sigma$, and 3.3-$\sigma$ for samples 1, 2, and 3 respectively.
Finally,
a Kolmogorov--Smirnov (KS) test was applied to $P(\phi \le \phi_{\rm MAX})$ 
for each of the samples and, again, in all cases an isotropic
distribution was rejected with high significance (see Fig.\ 2).

Fig.\ 3 shows the dependence of $\left< \phi \right>$ on the projected 
separation between the satellites and their hosts.  For a given sample, the
widths of the bins were chosen such that an equal number of satellites 
fell into each bin.  
Again, the dotted line indicates the expectation of
an isotropic distribution.
From Fig.~3 it is clear that the anisotropy in the distribution of the
satellites is strongest on relatively small scales ($r_p 
\lesssim 100$~kpc),
and on scales of $250~{\rm kpc} \lesssim r_p \lesssim 500~{\rm kpc}$  all three
samples are consistent with a uniform distribution at of order the 1-$\sigma$ level.
It is very unlikely that on scales $r_p \lesssim 100~{\rm kpc}$ 
a large number of giant H--II regions (i.e., analagous
to 30 Doradus) have been misidentified as ``satellites'' for a number of
reasons.  First, only objects for which the SDSS spectra are clearly classified
as ``galaxies'' have been used.  Also, the ratio of the luminosity of the hosts to
the luminosity 
to these objects is much smaller than would be expected for giant H--II regions 
(median
luminosity ratios of 13, 10, and 24 for samples 1, 2, and 3 respectively).  
Finally, the isophotal areas of these objects compared to their hosts is much
larger than would be expected for giant H-II regions (median isophotal area
ratios of 0.09, 0.12, and 0.04 for samples 1, 2, and 3 respectively).

%
\vspace*{0.2truecm}
\hbox{~}
\centerline{\psfig{file=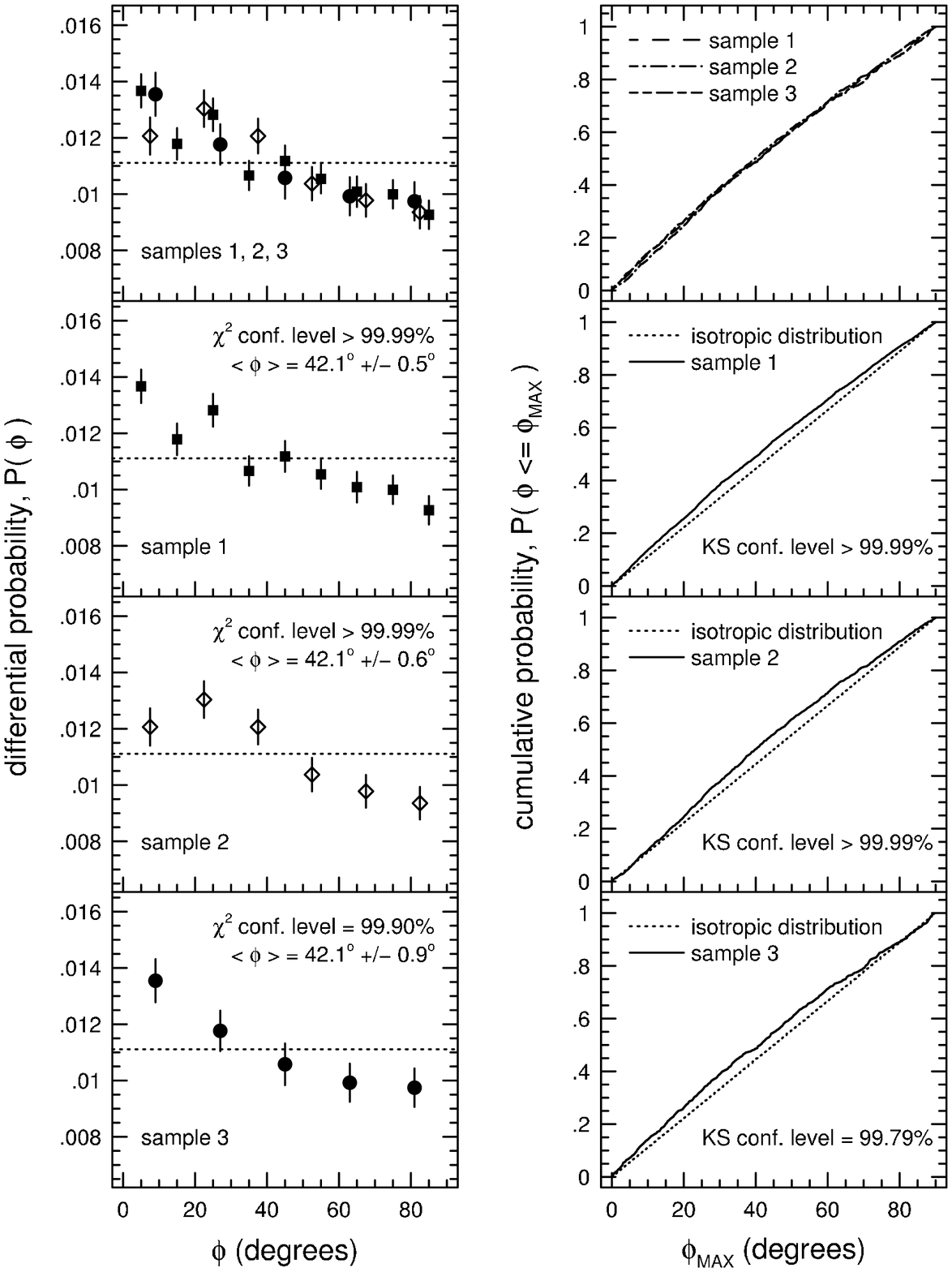,angle=0,width=3.25in}}
\vspace*{-0.7truecm}
\noindent{\scriptsize
Figure~2.
Left panels: Differential probability distribution
for the location of satellites relative to the host major axis.
Error bars were computed from 1000 bootstrap resamplings of the data.
Dotted line shows $P(\phi)$ for an isotropic distribution.
Formal rejection confidence levels from the $\chi^2$ test
are shown in the panels, along with $\left< \phi
\right>$ for each of the samples.
Right panels: Cumulative probability distribution
for the location of satellites relative to the host major axis.  Formal
rejection confidence levels from the KS test are shown in the panels.
\label{fig2}
}

\section{Discussion and Conclusions}

All three samples 
of satellites show a strong preference for being
aligned with the host major axis on small scales ($r_p 
\lesssim 100~{\rm kpc}$),
while on larger scales ($250~{\rm kpc} \lesssim r_p \lesssim 500~{\rm kpc}$)
the satellite distribution
is fairly consistent with an isotropic distribution.
These results
contrast sharply with most previous  
studies; i.e., when
an anisotropy in the distribution of satellites has been seen in the
past, the satellites have generally been aligned with the host
minor axis, not the host major axis.  
Aside from this {\it Letter},
the most recent investigation of the distribution of satellite galaxies
relative to host galaxies is that of SL, who 
concluded that the satellites of
2dFGRS host galaxies are preferentially aligned with the minor axes
of the hosts, but only for a very select subset of the systems.  That
is, SL selected hosts and satellites in exactly the same way in which 
sample 1 of this {\it Letter} was selected, and host--satellite pairs for which
the relative velocity difference is $|dv| < 500$~km~sec$^{-1}$ were
specifically included in SL's work.  When
SL examined their entire sample of satellites, no anisotropy in the 
distribution of satellites was found.  When they restricted the
sample to hosts and satellites with $|dv| < 160$~km~sec$^{-1}$,
a strong detection of the Holmberg effect was seen.  In particular,
SL report that
the ratio of the number of satellites found within
$30^\circ$ of the host major axis to the number of satellites
found within $30^\circ$ of the host minor axis is
$f = 0.80 \pm 0.04$ for hosts and satellites that have low relative
velocity.  In contrast, the data presented here
yield $f = 1.30 \pm 0.08$ for all of the
satellites in sample 1 and $f = 1.28 \pm 0.10$ for hosts and satellites
in sample 1 for which $|dv| < 160$~km~sec$^{-1}$.

%
%
\vspace*{0.2truecm}
\hbox{~}
\centerline{\psfig{file=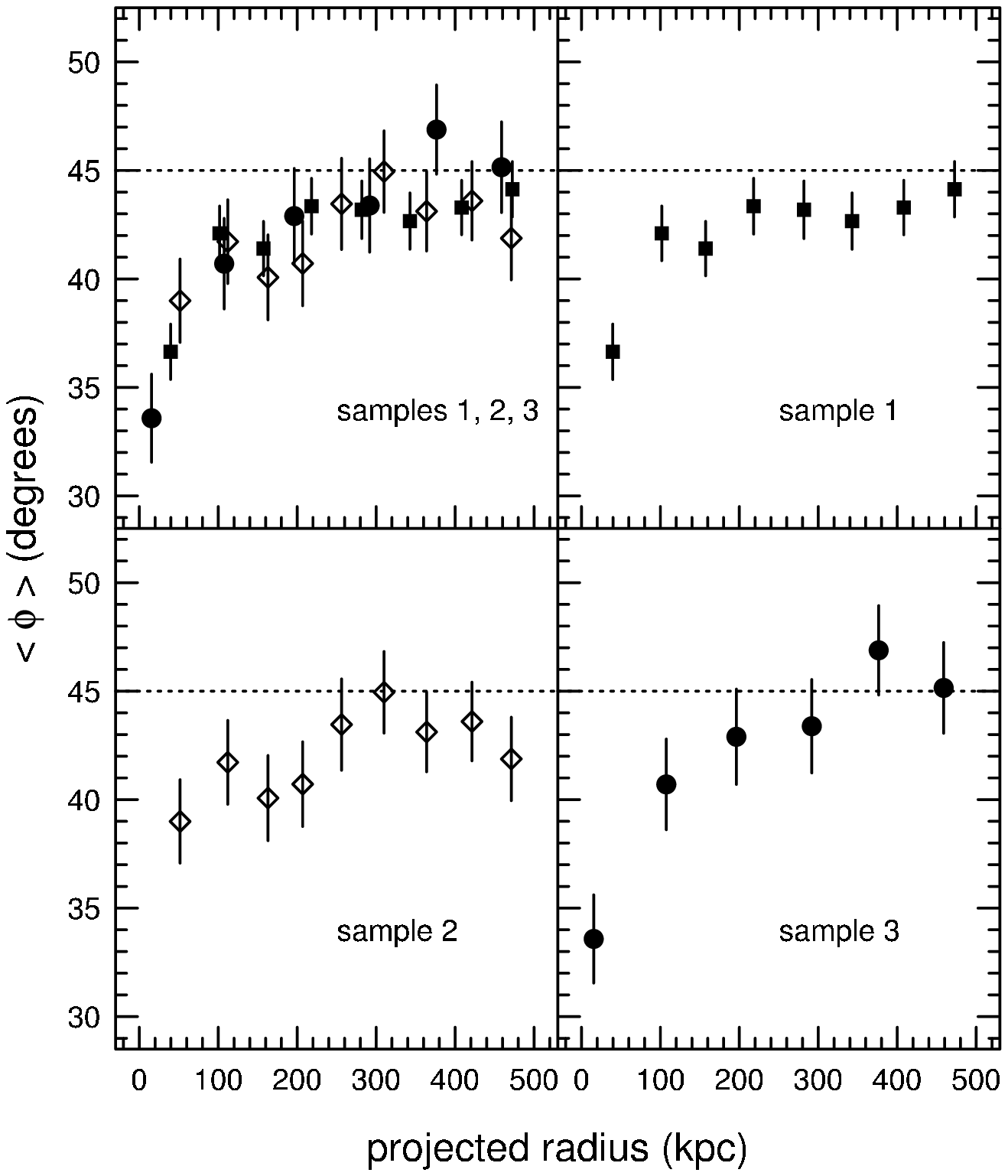,angle=0,width=3.0in}}
\noindent{\scriptsize
Figure~3.
Mean orientation of satellites with respect to the host major axis as
a function of the projected radius.
\label{fig3}
\vspace*{0.2truecm}
}

The SL result is troubling because the typical error
on a radial velocity measurement in the 2dFGRS is of order 85~km~sec$^{-1}$,
so a measured value of $|dv| = 160$~km~sec$^{-1}$ is comparable to the error in the
value of $|dv|$.  (In the case of the SDSS, the typical velocity errors
are of order 20 to 30~km~sec$^{-1}$.)  In addition,  simulations of
galaxy redshift surveys by van den Bosch et
al.\ (2004) show
that the fraction of interloper galaxies is higher for host--satellite
pairs with small values of $|dv|$ than it is for host--satellite pairs
with large values of $|dv|$.  It is possible, therefore, that the host--satellite
pairs in SL with $|dv| < 160$~km~sec$^{-1}$ are heavily contaminated by interlopers
and some unknown selection bias is responsible for the signal which they
detect.  Further, the results in this {\it Letter}, combined with the results
of van den Bosch et al.\ (2005), may provide some insight into the 
apparent lack of anisotropy seen by SL in their full sample.  Fig.\ 3
of this {\it Letter}
indicates that the anisotropy occurs primarily on  small 
physical scales.  However, when
van den Bosch et al.\ (2005) combined mock galaxy redshift
surveys  with the 2dFGRS, they found a marked absence of satellites at small
projected radii,
 which they attributed in part to the merging and overlap of images
in the APM catalog.  That is,
SL might have had too few 
host--satellite pairs with $r_p \lesssim 100$~kpc to detect the anisotropy 
that is seen here in the SDSS DR3 samples.

Following on from the numerical work of Knebe et al.\ (2004), who first
predicted a preferential alignment of satellites with the major axes of
primary galaxies in cluster environments, the most recent theoretical work
on the distribution of the satellites of isolated host galaxies
shows that in a CDM universe the satellites ought to be found preferentially
close to the principle plane of the host halo (i.e., the plane defined by
the two largest moments of inertia of the host halo). 
High--resolution simulations of the
formation of ``Milky Way'' systems show that the satellite galaxies form
a rather thin, planar structure that lies within the principle plane of
the halo (e.g., Libeskind et al.\ 2005; Zentner et al.\ 2005).   In addition,
AB showed
that, in projection on the sky, when hosts and satellites are 
selected from a $\Lambda$CDM simulation that incorporates semi-analytic
galaxy formation, the satellites show a strong preference for clustering
near to the major axes of the projected host halos.

AB placed circular disks in the principle planes of the host halos
(i.e., to represent the disks of the host galaxies since the simulation
did not contain images of visible galaxies) and they then computed the
distribution of satellites with respect to the major axes of the projected
circular disks using 100 random rotations of the simulation.  
AB found that, in the plane of the sky, satellites
with $r_p < 500$~kpc
showed a clear preference for clustering nearby to the major axes of the
projected circular disks and, in fact, the degree of anisotropy shown
by AB's satellites was
somewhat greater than the degree of anisotropy shown by the SDSS
satellites.  The anisotropy shown by AB's satellites is caused
by the fact that in a CDM universe satellites
are accreted preferentially into the principle plane of the host halo 
along filaments (see also Libeskind et al.\ 2005 and Zentner et al.\ 2005).

AB were, however, unable to reproduce the trends shown in Fig.\ 3; i.e., 
that the anisotropy in the distribution of the SDSS satellites 
arises primarily on scales $\lesssim 100$~kpc.  Contrary to the results
of this {\it Letter},
AB found that the anisotropy in the distribution of the simulated
satellites persisted to large scales ($r_p \sim 500$~kpc).    Since the
hosts and satellites in the simulation were selected in exactly the same
way as was done here for the SDSS, the fraction of galaxies in the SDSS
that may have been falsely identified as satellites (i.e.,  ``interloper galaxies'')
should be similar in both cases.  Hence, the well--known increase in the
number of interlopers with increasing $r_p$ (e.g., Prada et al.\ 2003;
Brainerd 2004) is unlikely to be the cause of the SDSS satellites showing
a nearly isotropic distribution at large $r_p$.  It is not clear
what the underlying cause of the discrepancy between the
degree of large--scale anisotropy shown by AB's satellites and the SDSS
satellites happens to be, but a likely possibility is that the disks
of host galaxies may be aligned only with the very innermost parts
of their halos (whereas AB used all mass within the virial radius to
define the plane in which the host's disk would reside).  Recent work by 
Bailin et al.\ (2005) on the formation of disk galaxies in a CDM universe
shows that
although the orientation of the disk is aligned well with the
halo on small scales, the orientation of the disk is essentially
uncorrelated with the outer regions of the halo.  Since the satellites
presumably trace the gravitational potential and accretion history
of the halo (i.e., the mass), rather than the disk specifically, Bailin
et al.'s result would lead directly to the expectation that on very large
physical scales the satellites
of disk galaxies should be distributed roughly isotropically relative
to the direction vector defined by the major axis of the disk.

In conclusion, the satellites of host galaxies in the SDSS DR3 show a
marked preference for alignment with the major axes of the hosts.
Although the sense of this anisotropic
distribution agrees well with the expectations
of a CDM universe, a proper comparison of the details of the signal
(i.e., a strongly anisotropic distribution on small scales and a
nearly isotropic distribution on 
large scales) awaits high resolution simulations that are capable of
fully addressing the orientations of visible galaxies within their
dark matter halos.

\section*{Acknowledgments}

Support under NSF contract AST-0406844
is gratefully acknowledged.
Funding for the SDSS has been provided by the 
Alfred P. Sloan Foundation, the Participating Institutions, NASA,
the NSF, 
the US Dept.\ of Energy, the Japanese Monbukagakusho, and the Max 
Planck Society. The SDSS Web site is http://www.sdss.org/.
The SDSS is managed by the Astrophysical Research Consortium for 
the Participating Institutions 
(Univ.\ of Chicago, Fermilab, the Institute for Advanced Study, the 
Japan Participation Group, Johns Hopkins Univ., Los Alamos National 
Laboratory, the Max-Planck-Institute for Astronomy, the 
Max-Planck-Institute for Astrophysics, New Mexico State Univ., 
Univ.\ of Pittsburgh, Princeton Univ., the US Naval 
Observatory, and Univ.\ of Washington).



\end{document}